   \documentclass[manuscript]{aastex}
\usepackage{graphicx}
\usepackage{epstopdf}

\shorttitle{Missing chromsospheric connection}
\shortauthors{M. Battaglia et al.}
\begin{document}
\renewcommand{\labelitemi}{-}
\title{WHERE IS THE CHROMOSPHERIC RESPONSE TO CONDUCTIVE ENERGY INPUT FROM A HOT PRE-FLARE CORONAL LOOP? }
\author{Marina Battaglia\altaffilmark{1}
  \and Lyndsay Fletcher\altaffilmark{2} \and Paulo J. A. Sim\~oes\altaffilmark{2}}
\affil{Institute of 4D Technologies, School of Engineering, University of Applied Sciences and Arts Northwestern Switzerland, CH-5210 Windisch, Switzerland} 
\affil{SUPA School of Physics \& Astronomy, University of Glasgow, Glasgow G12 8QQ, UK}

\email{marina.battaglia@fhnw.ch}

\begin{abstract}
Before the onset of a flare is observed in hard X-rays there is often a prolonged pre-flare or pre-heating phase with no detectable hard X-ray emission but pronounced soft X-ray emission suggesting that energy is being released and deposited into the corona and chromosphere already at this stage. 
This work analyses the temporal evolution of coronal source heating and the chromospheric response during this pre-heating phase to investigate the origin and nature of early energy release and transport during a solar flare.
Simultaneous X-ray, EUV, and microwave observations of a well observed flare with a prolonged pre-heating phase are analysed to study the time evolution of the thermal emission and to determine the onset of particle acceleration. During the 20 minutes duration of the pre-heating phase we find no hint of accelerated electrons, neither in hard X-rays nor in microwave emission. However, the total energy budget during the pre-heating phase suggests that energy must be supplied to the flaring loop to sustain the observed temperature and emission measure. Under the assumption of this energy being transported toward the chromosphere via thermal conduction, significant energy deposition at the chromosphere is expected. However, no detectable increase of the emission in the AIA wavelength channels sensitive to chromospheric temperatures is observed.
The observations suggest energy release and deposition in the flaring loop before the onset of particle acceleration, yet a model in which energy is conducted to the chromosphere and subsequent heating of the chromosphere is not supported by the observations. 
\end{abstract}
\maketitle
\section{INTRODUCTION} \label{Introduction}
In the commonly accepted flare model, acceleration of particles (electrons and protons) takes place in the corona, near or above the top of magnetic loops. Beams of fast electrons precipitate downward along the loop and their energy is deposited in the chromosphere, where hard X-ray (HXR) bremsstrahlung is produced, visible as chromospheric footpoints. The energy deposition in the chromosphere leads to heating and expansion of the chromospheric plasma into the loop, termed chromospheric evaporation. The hot plasma becomes visible as soft X-ray (SXR) coronal sources and EUV loops. In such a scenario one would expect a close association of the HXR lightcurves and the SXR lightcurve, as first suggested by \citet{Ne68} and indeed observed in a large percentage of flares \citep[e.g.][]{De93,McT99,Ve05}. However, some flares exhibit a prolonged phase of pre-impulsive activity during which the SXR flux increases steadily with time but with no observed HXR emission \citep{Ac92}. \citet{Ba09} analysed in detail a number of such pre-heating events observed with RHESSI \citep{Li02}. They see an increase in emission measure during the pre-heating phase which they attribute to an increase of the coronal loop density due to chromospheric evaporation caused by conductive heating of the chromosphere. Recently, \citet{2012ApJ...758..138A} re-analyzed two of these events, including radio observations. They find indication of a non-thermal electron population before an increase in HXR emission is observed, showing that acceleration might start before RHESSI is sensitive enough to see the associated HXR emission. However, the energy in these particles is not enough to explain the pre-heating. The authors further attribute the increase in emission measure to an increase of emitting volume, arguing that chromospheric evaporation is not necessary to explain the observations. But regardless of this, a connection between the coronal source and the chromosphere has to be expected. Energy transported to the chromosphere, be it by conduction or particle beams or any other means will change the chromospheric environment. This may not be observable in either X-rays nor microwaves but most likely in EUV. Recently \citet{2013ApJ...771..104F} analysed flaring ribbons before the onset of the impulsive phase where they clearly see emission in all wavelength channels of SDO/AIA \citep{Le11} and even suggest that the majority of GOES and RHESSI thermal emission in this early phase is of chromospheric origin. 

Here we present multi-wavelength observations of a flare with a prolonged ($\approx$ 20 minutes) pre-heating phase. RHESSI images and spectra are combined with SDO/AIA observations, as well as data from the Nobeyama Radioheliograph (NoRH; \citet{1994IEEEP..82..705N}) and Nobeyama Radio Polarimeter (NoRP; \citet{1985PASJ...37..163N}) to investigate the temporal evolution of the hot loop and the response of the chromosphere during the pre-heating phase. Hard X-ray and radio data suggest that there were no significant non-thermal particles present during the first 20 minutes of the flare. However, the emission measure increases continuously during this phase, indicating continuous heating of the coronal source. Assuming Spitzer conductivity, the total energy flux toward the chromosphere is calculated. In this scenario a large amount of energy is expected to be deposited at the chromosphere, yet we observe no significant increase in emission at 304 \AA , 1600 \AA , or 1700 \AA\  or in any of the other AIA EUV channels. This is in contradiction to the expectation and lacks an immediate explanation. In Section 2 we describe data analysis in the different wavelengths, in Section 3 the time evolution of the total flare energetics is presented and discussions and conclusions drawn in Sections 4 and 5.
\section{DATA ANALYSIS} \label{overview}
\begin{figure}[] 
\begin{center}
\includegraphics[height=15cm]{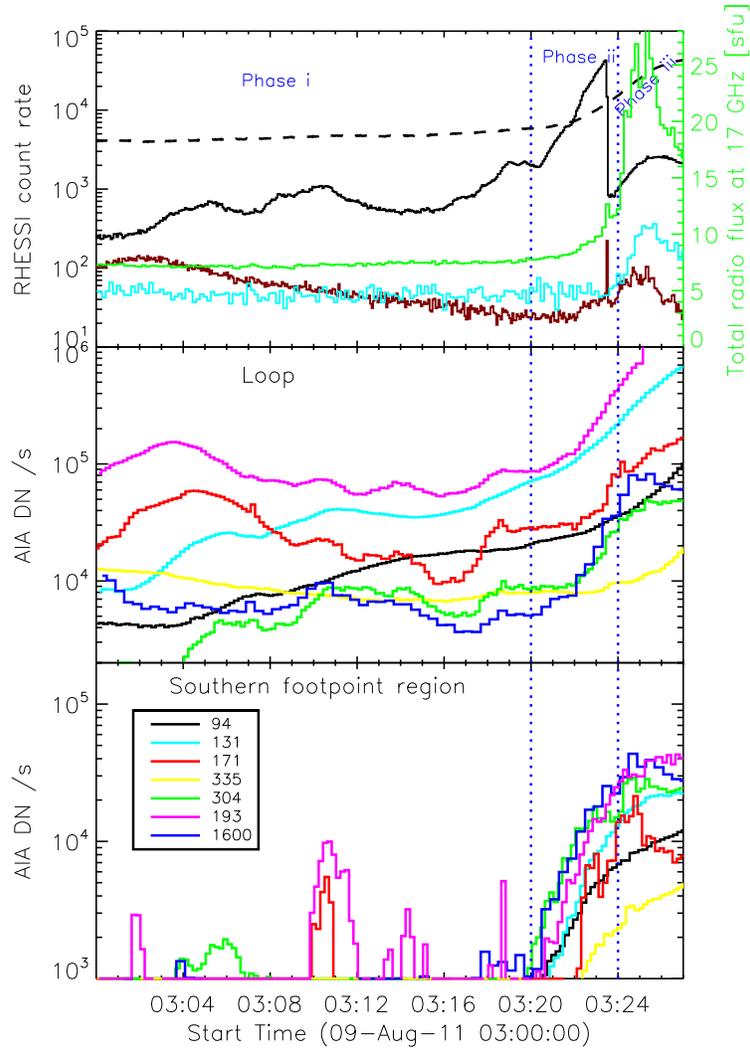}
\end{center}
\caption {Top: RHESSI 6-12 keV (black) and 25-40 keV (red) count rate lightcurves. GOES lightcurve (dashed line) and NoRH 17 GHz (green) and 34 GHz (blue) lightcurves. The RHESSI 25-40 keV lightcurve before 03:15 shows signatures of non-solar emission. Middle: Background-subtracted AIA lightcurves of the EUV loop in 7 wavelength channels (see legend). Bottom: Background-subtracted AIA lightcurves in 7 wavelength channels from the region of the southern footpoint. }
\label{partimevol}
\end{figure}

The flare (SOL2011-08-09T03:52) happened on August 9th 2011 with the first hard X-ray HXR peak at 03:25 UT. The soft X-ray SXR flux started to increase at around 03:00 UT, 25 minutes before the HXR peak. Note that the flare peak in the GOES light-curve was observed only 25 minutes afterward (at around 03:52 UT) but RHESSI went into night at 03:35 UT. Thus the whole time interval analyzed here can be viewed as pre-flaring activity and the main HXR peak was probably missed by RHESSI. Equally there was a short SXR burst at an earlier time (around 02:40 UT). Since the SXR flux returned to near background level before 03:00 UT and RHESSI images suggest that the location of this earlier peak was displaced by $\sim$ 10 arcsecs,  we chose the start-time of the time-interval of interest as 03:00 UT, when RHESSI imaging became possible. We divide the flare into three distinct phases according to the evolution of the X-ray lightcurves: (i) Pre-heating (until $\sim$ 03:20 UT). No noticeable HXR or microwave emission is observed during this phase, but the SXR emission gradually increases (ii) Onset of particle acceleration. The SXR flux increases by one order of magnitude and a tail toward higher X-ray energies becomes gradually visible in the spectrum (from $\sim$ 03:20 - 03:24 UT). (iii) HXR peak with peak time at around 03:25 UT. During this phase, HXR emission is observed up to 40 keV. In this work we focus on phase (i) and the transition to phase (ii).  Figure \ref{partimevol} (top) shows RHESSI lightcurves at 6-12 keV and 25-40 keV, GOES lightcurve, and the flare integrated Nobeyama 17 GHz and 34 GHz lightcurves, illustrating the three phases. 
\subsection{X-ray analysis} 
\begin{figure}[]
\begin{center}
\includegraphics[height=10cm]{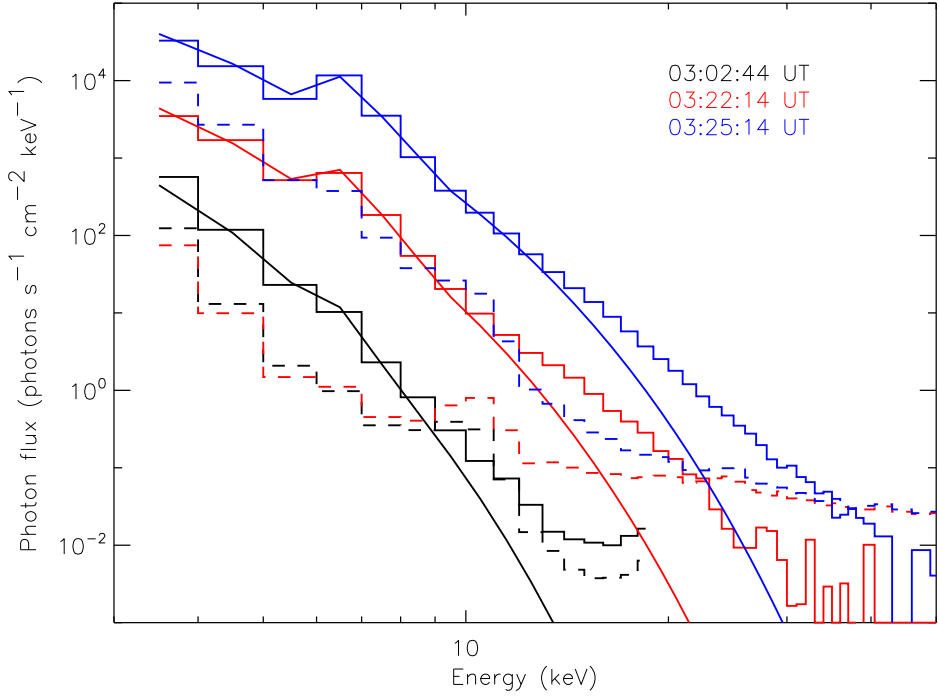}
\end{center}
\caption {Top: RHESSI spectra at three different times, corresponding to the three flare phases. The dashed lines give the background at the corresponding time. The thermal fit model only is indicated to better illustrate the appearance of a ``tail'' above the thermal emission toward higher energies later in the flare. The fitted temperature and emission measures at the three times were 10 MK / $1.7\times 10^{46}$ cm$^{-3}$, 15.9 MK / $6.8\times 10^{46}$ cm$^{-3}$, and 20 MK / $4.7\times 10^{47}$ cm$^{-3}$. The spectrum at 03:02:28 UT was dominated by non-solar emission at energies larger than $\sim$ 15 keV (compare Figure \ref{partimevol}) and is not shown. }
\label{spectrum}
\end{figure}

RHESSI spectra were fitted using OSPEX starting from 03:00 UT over time intervals of $\sim$ 30 seconds. The spectrum was fitted with a single thermal component giving the temperature and emission measure up until 03:22 UT, when a tail to higher energies started to appear. This tail was fitted with a soft (photon spectral index $\gamma \ge 6.5$) non-thermal power-law component. Note that at this time the live-time of the RHESSI detectors was $\sim$ 90 \% and it is possible that up to 50 \% of the high energy flux consisted of pile-up counts \citep{Smith02}. Indeed, fitting a pile-up component as provided by OSPEX leads to a smaller non-thermal flux by one order of magnitude and to a steeper spectral index of $\gamma=7.1$. Thus, the fitted non-thermal component has to be viewed as an upper limit during this phase. Figure \ref{spectrum} shows the RHESSI spectrum near the start of the pre-heating (phase i), the first appearance of the high-energy tail (phase ii), and the HXR peak (phase iii). In addition to the RHESSI thermal parameters the background-subtracted GOES emission measure and temperature were determined using the GOES utilities in SSW. However, the GOES emission measure could only be determined reliably after 03:20 UT, possibly due to the limited sensitivity of GOES and low counts prior to this time. 

\begin{figure*}[]
\begin{center}
\includegraphics[height=14cm]{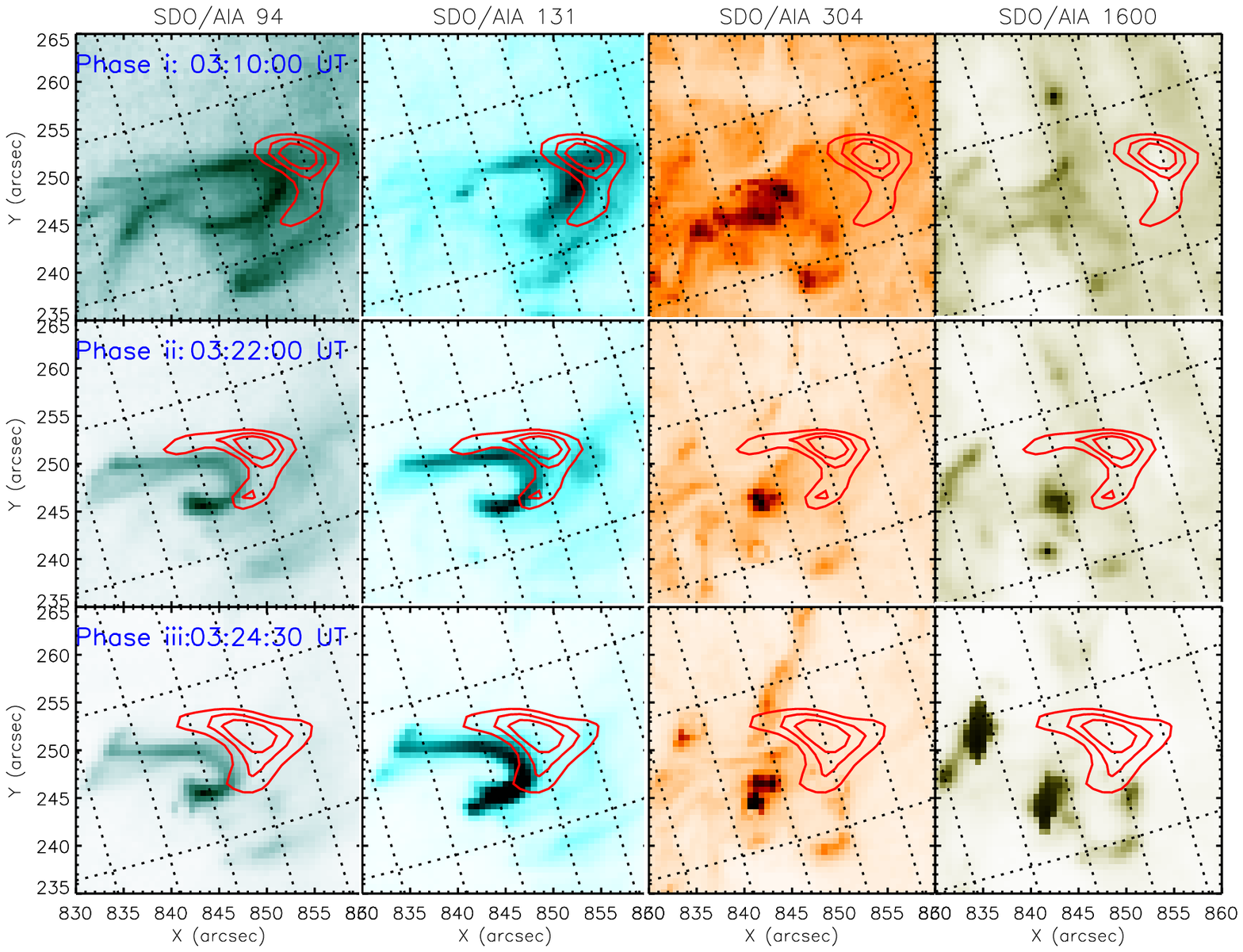}
\end{center}
\caption {AIA images in 4 wavelength channels at 3 different times corresponding to phases i - iii. 30, 50, and 70 \% contours from RHESSI MEM\_NJIT 6-12 keV images taken over $\approx$ 1 minute intervals around the time of the AIA image are overlaid. }
\label{aiarhoverlay}
\end{figure*}

RHESSI imaging was possible starting from 03:00 UT. CLEAN images show the presence of a single X-ray source at all times and energies, i.e. the 6-12 keV source is co-spatial with the 25-50 keV source and no HXR footpoints are observed. In MEM\_NJIT images \citep{Sc07} which typically are more sensitive to smaller scale structures the outline of a loop similar to the loops observed in AIA's 131 \AA\ and 94 \AA\ wavelength channels is visible. This is illustrated in Figure \ref{aiarhoverlay}. The figure shows AIA images in four wavelength channels (94, 131, 304, and 1600 \AA\ ) at three representative times  ($\sim$ 03:10  = phase i; $\sim$ 03:22 UT = phase ii; $\sim$ 03:24:30 UT = phase iii)  overlaid with contours of MEM\_NJIT images taken around the times of the respective AIA images.
\subsection{EUV analysis}
\begin{figure}[]
\begin{center}
\includegraphics[height=16cm]{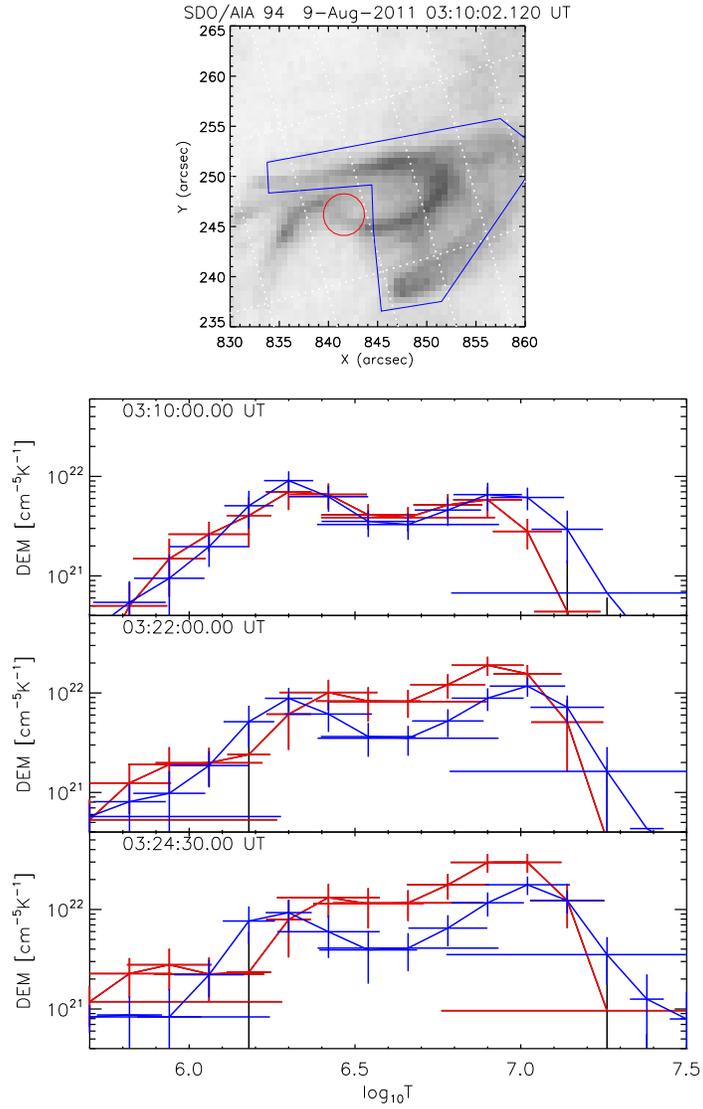}
\end{center}
\caption {Top: AIA 131 \AA\ image. The blue and red boxes indicate the regions of interest for which the AIA DEM was calculated. Other panels: AIA DEM per unit area from the two regions of interest for three representative times. }
\label{aiadem}
\end{figure}

SDO/AIA provides high spatial resolution images at 12 s cadence and in 9 wavelength channels covering temperatures from $\sim$ 5000 K up to $\sim$ 16 MK. Thus AIA complements RHESSI observations in terms of temperature sensitivity since RHESSI is sensitive to the hottest temperatures of a plasma (above $\sim$ 8 MK) and can often be fitted with a single temperature component even in the presense of cooler plasma \citep[e.g.][]{2012ApJ...760..142B, 2014SoPh..tmp...31R}. AIA images and lightcurves at different wavelengths give a good overview of the flare morphology and the timing and locations of predominant emission at a given temperature. In this study we distinguish between two regions of interest, namely the whole EUV loop and the location where the southern footpoint appears during phase (ii) (see Figures \ref{aiarhoverlay} and \ref{aiadem}). Background-subtracted lightcurves from each region are presented in Figure \ref{partimevol} for comparison with the RHESSI lightcurves where the background time was taken at 02:30 UT.
Beyond lightcurves, AIA data is also used for differential emission measure (DEM) analysis to investigate the temporal evolution of the emission measure and, ultimately, the density of the loop using an independent method from the RHESSI analysis. Here we use the regularization method developed by \citet{2012A&A...539A.146H}. The DEM was calculated for the two regions of interest. Representative examples for the three phases are shown in Figure \ref{aiadem}. The DEM shows two peaks in temperature, one at around 2 MK and one at around 10.2 MK. The bulk of the low temperature component can be attributed to foreground line of sight emission while the high temperature component is dominated by flaring emission \citep[e.g][]{2012ApJ...760..142B}. We define a total emission measure by integrating the DEM over temperature $EM_{AIA}=\int DEM dT$, where we integrated between 4.5 MK and 10.5 MK and between 10.5 MK and 18 MK (see Figure \ref{ethermal}). 
\subsection{Microwave analysis}
RHESSI is a high background instrument and non-thermal HXR emission has to reach quite high intensity for RHESSI to be observed. It has thus been argued \citep[e.g.][]{2012ApJ...758..138A} that particle acceleration may start much earlier but the HXR emission is too faint to be observed by RHESSI. Microwave observations give an independent measurement of non-thermal particles and thus help constrain the onset of particle acceleration. Radio observations from Nobeyama at 17 GHz and 34 GHz were available during the whole duration of the flare presented here. Figure \ref{partimevol} shows the lightcurve at 17 GHz. Comparison between RHESSI 25-40 keV lightcurves and Nobeyama 17 GHz lightcurves suggests a temporal correlation between the two emissions with the rise of the radio emission being simultaneous with the rise of the HXR emission. There was no significant increase of microwave emission before phase (iii) suggesting that particle acceleration was indeed minimal during the pre-flare phase. This is supported by an indepth analysis of NoRP spectra. We derived pre-flare average flux densities and their 1-$\sigma$ variations within the time interval 02:40~UT to 02:55~UT. None of the NoRP frequency channels show any sustained increase above their 1-$\sigma$ levels during phase (i). Over the same period, images at 17 and 34 GHz reveal stable emissions from the active region. Taking the maximum brightness temperatures $T_b$ of each frame, we found the average $T_b$ during phase (i) to be $2 \pm 0.1 \times 10^5$~K $2 \pm 0.3 \times 10^4$~K at 17 and 34 GHz, respectively. Such values can be reasonable explained by optically thin free-free emission from a thermal plasma at 2~MK (or even hotter), with a density of $\approx 1.2 \times 10^{10}$~cm$^{-3}$, in agreement with our SXR and EUV analysis. The spectral index $\log(T_b(17GHz) / T_b(34GHz))/\log(17/34)=-3.3$ is steeper than the theoretical value of $-2$ expected for an isothermal source, and this can be explained by a contribution of gyroresonant emission at 17 GHz \citep{1985ARA&A..23..169D}.  
\section{EVOLUTION OF FLARE ENERGETICS AND CHROMOSPHERIC RESPONSE} 
The RHESSI SXR lightcurve (fitted temperature $\sim$ 11 MK) and the AIA 131 \AA\ loop-lightcurve (peak in temperature response at $\sim$ 12 MK) exhibit a similar pattern during phase (i). The 94 \AA\ loop-lightcurve (peak in temperature response at $\sim$ 8 MK) also shows an increase in intensity during phase (i), but is not as well correlated with the RHESSI lightcurve. This suggests that plasma is heated during this phase and the bulk of the emitting plasma has a temperature of around 10-12 MK. Under the assumption of thermal conduction along the loop one would therefore expect chromospheric heating and chromospheric evaporation as argued by e.g. \citet{Za88,Ba09}. Using the temperature, emission measure, and the non-thermal parameters found from RHESSI and combining them with observations from SDO/AIA it is possible to investigate the total energy budget during the pre-heating phase and challenge the picture of chromospheric evaporation by looking for signatures in the AIA wavelength-bands that are sensitive to chromospheric temperatures. In a first step we investigate the temporal evolution of the thermal plasma parameters and give an estimate of the density. In a subsequent step the total thermal energy in the flare and the expected losses by thermal conduction and radiation are calculated. 
\subsection{Time evolution of thermal parameters} \label{sthermal}
\begin{figure*}[]
\begin{center}
\resizebox{0.95\hsize}{!}{\includegraphics{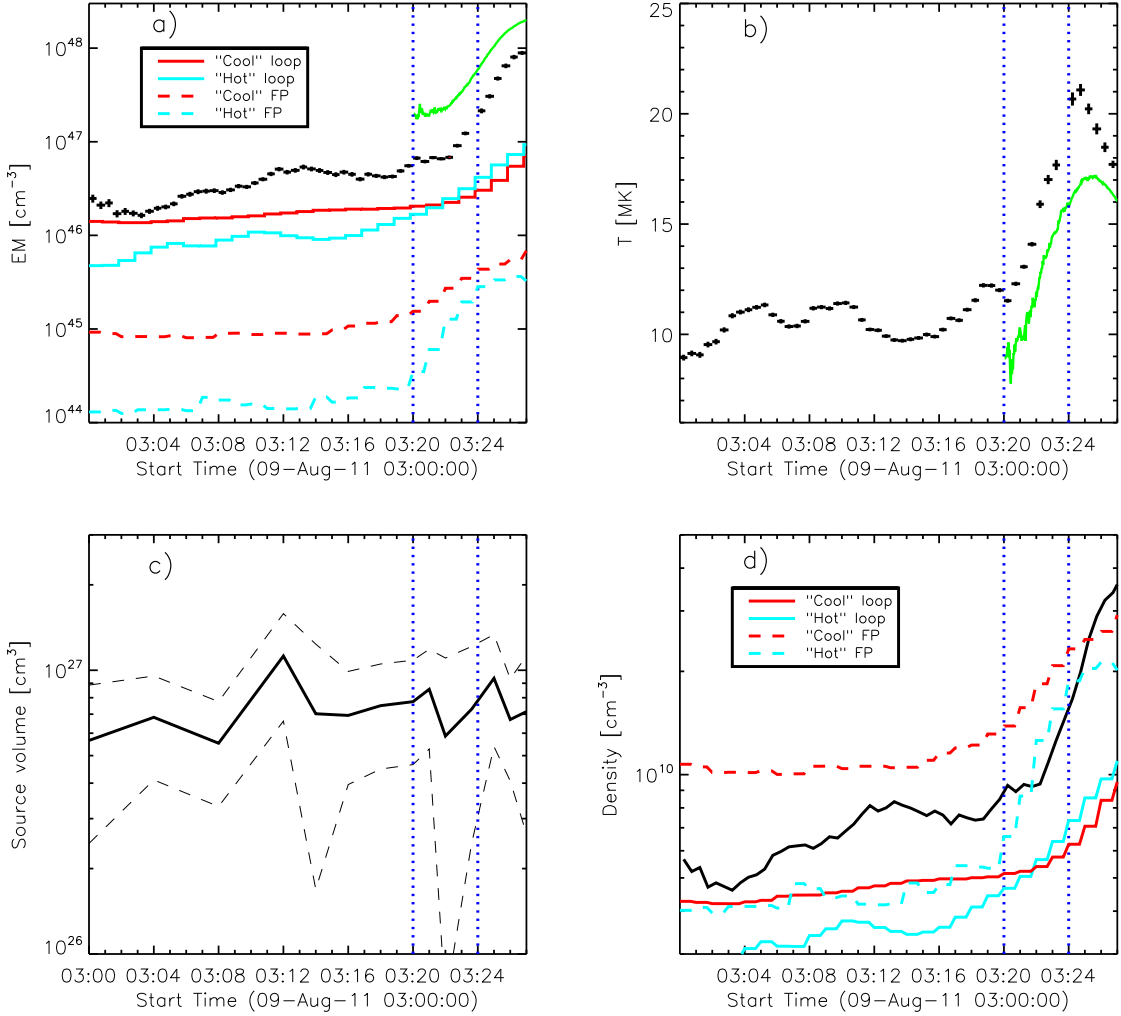}}
\end{center}
\caption {a) Time-evolution of total emission measure. The black points indicate RHESSI measurements, the green line gives the GOES values. The red line is the AIA emission measure integrated between 4.5 and 10.5 MK, the blue line gives the AIA emission measure integrated between 10.5 and 18.5 MK. b) RHESSI (black) and GOES (green) temperatures. c) RHESSI source size averaged from three imaging algorithms. The dashed lines give the confidence range. d) RHESSI density (black line). The red lines give the density from the total AIA emission measure between 4.5 and 10.5 MK of the EUV loop (solid line) and the footpoint region (dashed line).  The blue lines give the density from the total AIA emission measure between 10.5 and 18.5 MK of the EUV loop (solid line) and the footpoint region (dashed line).}
\label{ethermal}
\end{figure*}

Figure~\ref{ethermal}a gives the time evolution of the RHESSI emission measure and temperature, the GOES emission measure and temperature, and the AIA emission measure integrated over two temperature ranges (4.5 - 10.5 MK, 10.5 - 18 MK). 
The RHESSI emission measure during phase (i) shows an overall increase of about a factor of 3. Fluctuations in the RHESSI temperature in the pre-heating phase correlate with fluctuations in the 6-12 keV lightcurve but overall the temperature does not increase significantly. The total AIA emission measure of the EUV loop in both temperature ranges increases continuously during phase (i), while the emission measure of the footpoint region remains constant. 
From the RHESSI images the volume of the flaring loop can be estimated. We combine images made with the CLEAN and MEM\_NJIT algorithms with results from visibility forward fitting to get an estimate of the source volume with realistic uncertainties. For CLEAN and MEM\_NJIT the 50 \% contours in each image were used as an estimate of the source area. Visibility forward fitting gives the FWHM of the fitted 2D Gaussian source from which the area can be calculated. The source area used in the further analysis of the event is the average of the three values with the $ 1 \sigma$ standard deviation giving an estimate of the uncertainty. The flare volume is then approximated as $V=A^{3/2}$ and the loop density $n_{loop}$ calculated via the fitted emission measure as $n_{loop}=\sqrt{EM/V_m}$, where we used the time-averaged source volume $V_m=(7.7 \pm 4.0) \times 10^{26}$ cm$^3$ since, as indicated in Figure \ref{ethermal}, the loop-volume did not show a significant increase during the pre-heating phase. This suggests that the observed increase in emission measure is not due to an increase in emitting volume as suggested by \citet{2012ApJ...758..138A}. The footpoint density can be estimated in a similar fashion using the AIA emission measure. Assuming a footpoint area of $A_{fp}=3.7\times 10^{16}$ cm$^{2}$ (about the size of the 304 \AA\ footpoint visible at $\sim$ 03:21:56, compare Figure \ref{aiarhoverlay}), the footpoint density is then $n_{fp}=\sqrt{EM/(A_{fp}\times l_s)}$ where the line of sight component $l_s$ was taken as 3 arcsec (the approximate thickness of the chromosphere).
\subsection{Energy budget}\label{sebudget}
\begin{figure}[]
\begin{center}
\resizebox{0.95\hsize}{!}{\includegraphics{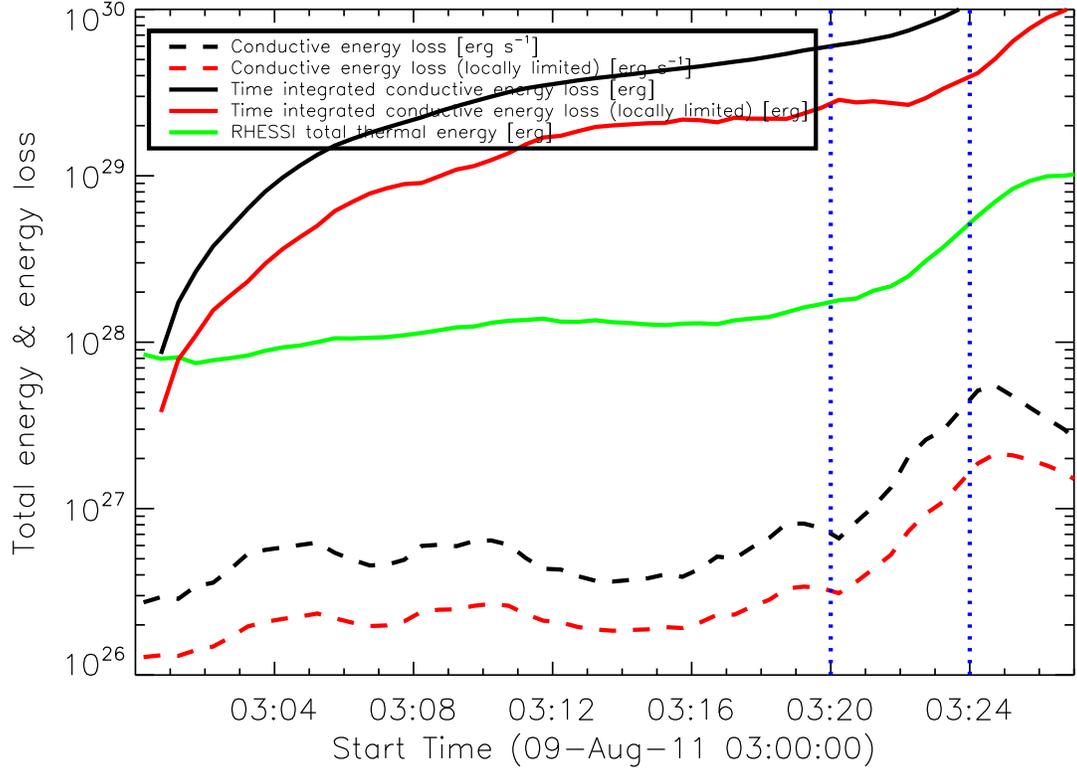}}
\end{center}
\caption {Time evolution of total thermal energy from RHESSI (green line), conductive losses in [erg s$^{-1}$ cm$^{-2}$] (dashed lines) and time-integrated (solid lines) in the case of classical Spitzer conductivity (black) and with locally limited conductive flux (red)} 
\label{condloss}
\end{figure}

The time evolution of the lightcurves and the emission measure of both RHESSI and AIA suggests continuous energy input during the pre-heating phase. Also, since we chose the start of the analysis interval as the time when RHESSI imaging became possible, a significant amount of thermal energy is already present at the start, as suggested by Figures \ref{ethermal} and \ref{condloss}. It is possible that some of the hot plasma present at 03:00 UT stems from the earlier flare at 02:40 UT (see Section 2). However, this only affects the start values of the density and emission measure but not the subsequent evolution. We therefore investigate the total thermal energy budget, including losses due to thermal conduction and radiation starting from 03:00 UT. The total thermal energy contained in the loop plasma with temperature T and emission measure EM at time $t$ is given as:
\begin{equation}
E_{th}=3k_BT\sqrt{EM\times V}\,\mathrm{[erg]}
\end{equation}
where $k_B$ is the Boltzman constant. The conductive losses in classical Spitzer conductivity \citep{Spbook} are given as:
\begin{equation} 
L_{cond}=10^{-6}\frac{T^{7/2}}{l_{loop}}\, \mathrm{[erg/s/cm^2]}
\end{equation}
where $l_{loop}$ is the length of one loop leg. The total loss through a loop with footpoint area $A$ is then $L_{cond}^{tot}=A\times L_{cond}$ erg s$^{-1}$.
However, for sufficiently large temperature gradients, such as occur in solar flares, the heat flux is expected to saturate, where two regimes of flux saturation can be defined \citep{Gra80}. \citet{Ba09} showed that for typical flare densities and temperatures, including the conditions in the flare presented here, a regime is reached where the conductive heat flux is locally limited. This can be accounted for by multiplying the classical conductive flux with a reduction factor $\theta=A\times exp(-b(\ln\mathcal{R}+c)^2)$ with A=1.01, b=0.05, c=6.63 and $\mathcal{R}=\lambda_{emf}/L_{th}$ where $\lambda_{emf}$ is the electron mean free path and $L_{th}$ the temperature scale length \citep{Ca84,Ba09}.
The time evolution of the total thermal energy and the conductive losses is shown in Figure \ref{condloss} where we assumed a footpoint area of $A=3.7\times 10^{16}$ cm${^2}$ (compare Section \ref{sthermal}) and a loop half-length of $l_{loop}=5.8\times 10^{8}$ cm, estimated from AIA 131 \AA\ images. 
The coronal radiative losses are several orders of magnitude smaller than the conductive losses during pre-flare conditions and can therefore be neglected.
The figure suggests that the time integrated conductive flux, even in the case of locally limited conductivity, is larger than the total thermal energy content and that energy of the order of $10^{10}$ erg cm$^{-2}$ is deposited into the chromosphere at each second. We now compare this with an estimate of the energy that is radiated away from the chromosphere. Since the thermal parameters and the DEMs are changing slowly during phase (i) we assume an equilibrium state between input energy and radiated energy and investigate one representative time-interval. The radiated energy from the footpoints per unit volume at a given temperature $T$ can be written as
\begin{equation}
E_R(T)=\Lambda(T)\times \frac{EM}{V}=\Lambda(T) \times n_e^2 \,\mathrm{[erg\,cm^{-3}s^{-1}]}
\end{equation} 
where $\Lambda(T)$ is the radiative loss function. This is related to the DEM as: 
\begin{eqnarray}
E_R(T)\frac{dh}{dT}&=&n_e^2\frac{dh}{dT}\Lambda(T) \\
E_R(T)\frac{dh}{dT}dT&=&DEM(T)\Lambda(T)dT \\
\end{eqnarray}
 with $DEM(T)=n_e^2dh/dT$. Assuming that all energy that is input by conduction is radiated away, then the radiated energy per unit area $A$ is:
\begin{equation}
\int E_R(T)dh=\int_{T_{min}}^{T_{max}} DEM(T)\Lambda(T)dT=\frac{F_{cond}}{A}
\end{equation}
Using the observed footpoint DEM at 03:10 UT (compare Figure~\ref{aiadem}) and the radiative loss function from CHIANTI for coronal abundances at a density of $10^{10}$ cm$^{-3}$, available through SSW, we get a radiated energy per unit area of the order of $6\times 10^6$erg $\mathrm{cm^{-2}s^{-1}}$, where we integrated over the whole temperature range for which the DEM was calculated (log$T_{min}=5.7$, log$T_{max}=7.5$). The conductive energy input, assuming a symmetrical loop with two footpoints of area $A=3.7\times 10^{16}$ cm${^2}$ would be $E_{cond} \sim 3.4\times 10^{9}$ erg cm$^{-2}$ s$^{-1}$, or, in the case of classical Spitzer conductivity, $E_{cond} \sim 8.0 \times 10^{9}$ erg cm$^{-2}$ s$^{-1}$. Thus the conductive energy input is expected to be a factor of between 600 and 1400 times larger than what is observed to be radiated away in the EUV between $10^{5.7}$ and $10^{7.5}$K. We note that the assumption of conductive flux driving radiative losses from the transition region means that we should narrow our range of temperature integration to that appropriate to the transition region, rather than this whole range which includes coronal temperatures. However, since it is not possible obtain a good transition region DEM from the AIA data, or to determine the exact boundary between transition region and corona from the observations, integration over the whole temperature range was done and the resulting radiative losses have to be viewed as upper limits. Note also that the DEM may contain emission from the foreground corona, particularly at lower temperatures up to $logT \sim 6.5$ \citep[see][for an extended discussion]{2012ApJ...760..142B}. Under this assumption the radiative losses would be a factor of 2 smaller, leading to an even larger discrepancy with the input energy by conduction. On the other hand the conductive losses depend on the loop length. Another method to estimate the loop length is to measure the footpoint separation and assume a semi-circular loop. In the present case, this results in a half-length of $l_{loop}=1.45\times 10^9$ cm, which is a factor of 2.5 more than our initial estimate. Thus the conductive losses could be reduced by about the same factor. 
\section{DISCUSSION}
The total energy calculation in the previous section suggests release of large amounts of energy during the pre-heating phase and subsequent heating of the loop. In a Spitzer model of thermal conduction one would expect conductive heating of the chromosphere and subsequent evaporation. This would result in increasing emission measure of the coronal loop and, if the loop volume remains constant, increasing density as observed by \citet{Ba09}. The flare presented here displays a prolonged pre-heating phase with increasing emission measure at a (constant) temperature of $\sim 10^7$ K. A conductive flux of the order of $5\times 10^{26}$ erg s$^{-1}$ suggests large energy deposition into the chromosphere but the estimate of the radiated energy based on the observed footpoint differential emission measure suggests that little energy does reaches the chromosphere. This is puzzling and lacks an immediate explanation. Observational uncertainties and the limitations of the available data and methods can only account for factors of 2-3 in both the calculated conductive energy input as well as the radiated energy whereas the observed discrepancy is a factor of at least a few hundred. 
Possible physical explanations include 1) Absorption of the observed EUV wavelengths by low-lying cool structures; 2) The energy is conducted elsewhere in the active region; 3) Inhibited conduction by trapping or ``bottling up'' of the electrons in the loop 4) Narrow, tenuous optically-thin radiating part of the plasma so that the bulk of the heat is deposited instead into the deeper, denser chromosphere where the energy is radiated at wavelengths we are not observing. In the following we dicuss these four explanations in detail. 

The first possibility is absorption by photo-ionization shortward of the Lyman edge at 912 \AA, by neutral hydrogen or neutral/once-ionised helium present in low-lying cool structures - typically the dark, cool inclusions in the transition region `moss' \citep[e.g.][]{2009ApJ...702.1016D}. In this case the observed footpoint DEM would be reduced and we would be underestimating the radiated energy in Equations 5-7. The analysis of \citet{2009ApJ...702.1016D} shows that for vertical viewing the absorption of EUV emission results in a decrease of a factor two, with a further reduction, thought to be a geometrical `line of sight' effect varying as the cosine of the source position angle. For the flare under study here, the position angle is roughly 60 degrees, giving an additional factor two of absorption, or a factor four overall. This would reduce the measured DEM by about the same factor and thus the emitted energy would still be two orders of magnitude smaller than the conducted energy. An increased absorption would be possible if the absorbing material were cooler or denser than typical moss; an increase of 5-6 in the optical depth would be sufficient. However, the footpoint densities that we find are consistent with moss plasma elsewhere on the Sun. The temperature can change the balance of absorbers (H I/He I/ He II) in the absorbing material, but to investigate this would require modeling that is outside the scope of this paper. We note that even if such absorption were present, one would still expect some increase of emission even from the absorbed lines, which is not seen. Furthermore, absorption of this kind could not explain why there is no significant change in the 1600 \AA emission, longward of the 912 \AA\ edge.

\begin{figure}[]
\begin{center}
\includegraphics[height=9cm]{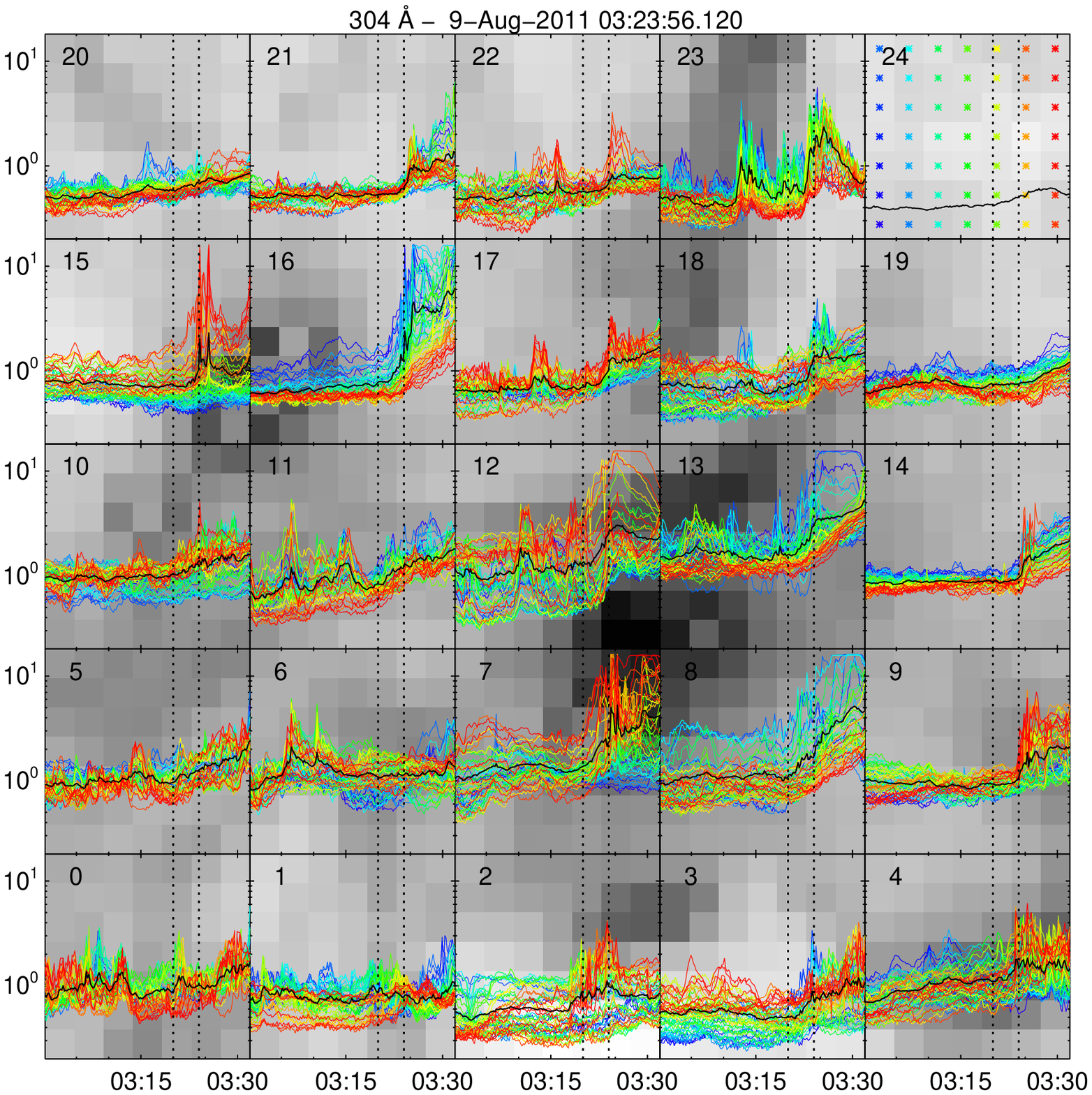} \\
\includegraphics[height=9cm]{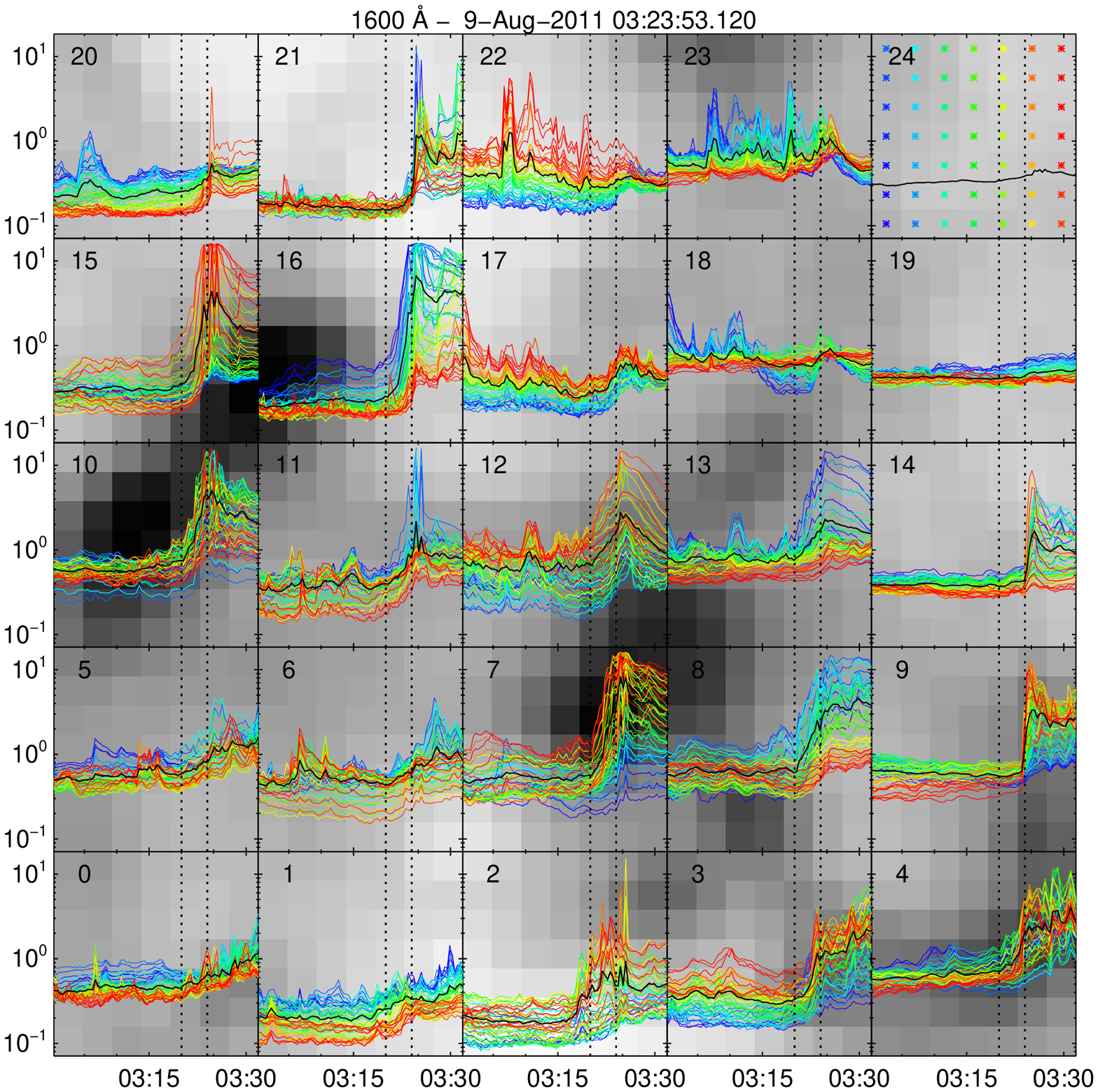} \\
\end{center}
\caption {AIA images at 304 \AA\ and 1600 \AA\ at $\sim$ 03:24 UT showing the loop footpoints overlaid with a grid of 25 squares. The time-evolution of the total flux within each square is given in black (normalized to the maximum in each square). The color coded lines (see top right square for the color key) give the pixel-by-pixel flux within each square. The vertical dotted lines give the end times of phases (i) and (ii). }
\label{grid}
\end{figure}

The second possibility is that the magnetic connections between the coronal source and the chromosphere are much more diverse than in our simple model with one magnetic loop and two footpoints. Indeed, AIA images suggest that there are multiple, fainter loops present during the whole pre-heating phase. It is therefore feasible that the energy is conducted to many different locations. We investigate this possbibility by analysing the time-evolution of the emission at different locations around the main flaring loop in several AIA wavelengths. We define a grid of 25 squares and calculate the time-evolution of the total flux within each square. Figure \ref{grid} shows AIA images at 304 \AA\ and 1600 \AA\  at $\sim$ 03:23:53 UT (flare peak) to illustrate where the footpoints appear during the main phase. The time evolution of the total flux within each square is overlaid. In addition, within each square the time evolution of the flux of each pixel is shown. The figure suggests that there are random brightenings in some regions during phase (i) but there is no correlation between any of these brightenings and the RHESSI or GOES lightcurves, neither for the square-averaged flux, nor the pixel-by-pixel flux evolution. Thus, while some energy could be conducted to ``elsewhere'' near the main flare site, the amount would not be enough to account for the total conducted flux. One could also consider a situation where the bulk of the energy is conducted away to the solar wind on open field-lines. However, the magnetic connectivity of the observed active region does not support such an interpretation. 
The third possible explanation is the existence of a process inhibiting thermal conduction so that it is significantly less than either the classical Spitzer rate or the locally limited rate described earlier in the paper. One possibility for this, explored by \cite{1979ApJ...228..592B}, is the `bottling up' of hot electrons by an ion-sound turbulent front, arising from the ion-sound turbulence generated when a hot electron population streams into a cooler ion population. The turbulence collisionlessly scatters the hot electrons, inhibiting their progress along the field, and reducing the rate of conduction. First of all we note that the electron and ion populations in the loop can be at different temperature. The electron temperature T$\sim 11$~MK and density at this temperature of $n_e \sim 5 \times 10^{9}$ cm$^{-3}$ mean that the electron-electron collision timescale $\tau_{ee}$ is short, and the electrons form a collisionally relaxed distribution.  \cite[][ p. 49]{2002ASSL..279.....B} gives $\tau_{ee} \sim {0.267 T^{3/2}}/{(\Lambda\ n_e)}$ where $\Lambda$ is the Coulomb logarithm, $\Lambda = \ln\; \bigl({8 \times 10^6 T /\surd{n_e}}\bigr)$. In the coronal loop $\Lambda = 20.9$, so $\tau_{ee} \sim 0.1$s  which is short compared to the evolution timescale of the source. The timescale for protons to equilibriate with the electrons is $m_p/m_e$ times longer, or $\sim 170$s. So at a given location in the loop there will be an interval of some perhaps 100s during which the electrons are hotter than the ions, and the condition for generation of the ion-acoustic turbulence is valid. According to \cite{1979ApJ...228..592B} a hot electron population in a loop will produce two narrow ion-acoustic turbulent fronts expanding along the loop at a speed $v_F$ equal to the local ion sound-speed
\begin{equation}
v_F = c_s = \biggl(\frac{kT_e}{m_p}\biggr)^{1/2}
\end{equation}
giving $v_F \sim 300\;{\rm km~s^{-1}}$, or $\sim 0.4"$ per second. The hot source would therefore expand to fill the estimated loop half-length 
$l_{loop}=5.8\times 10^8$~cm in $\sim 20s$, much shorter than the duration of the observations. The analytical results of \cite{1979ApJ...228..592B} show that the hot electron confinement time is thus $(m_p/m_e)^{1/2}$ times longer than the free-streaming time of the thermal electrons out of the loop, a result found also in the numerical experiments of \cite{2009ApJ...690..238A}. This can be compared to the conduction timescale given by classical Spitzer conductivity \citep[e.g.][p700]{2004psci.book.....A};
\begin{equation}
\tau_{\rm Spitz}=\frac{21}{5}n_e\frac{kL^2l}{\kappa T_e^{5/2}}
\end{equation}
evaluating this expression gives the ratio of the inhibited conduction timescale compared to the Spitzer conduction timescale:
\begin{equation}
\frac{\tau_{IA}}{\tau_{\rm Spitz}} = 1.9\times 10^5 \frac{T_e^2}{n_e L} = 7.9
\end{equation}
for the observed parameters of the source. The conductive flux is reduced by the same factor, compared to the classical Spitzer conductive flux. Observationally we require conduction to be limited by a factor $\sim 10^3$ (see Section \ref{sebudget}) compared to what classical conductivity would predict. We note also that towards the bottom of the loop the ion temperature drops (and the electron-ion temperature ratio increases) which will affect the generation of ion-acoustic turbulence, but this was also studied by \cite{2009ApJ...690..238A} who found only a very weak dependence of confinement time on ${T_e/T_i}$. Therefore bottling of the hot source by ion-sound turbulence as proposed by \cite{1979ApJ...228..592B} is insufficient to explain the observations. 
Another mechanism to confine electrons was proposed by \cite{1988ApJ...330..997S} in the form of an electrostatic trap. However, this only works for non-collisional electrons and is therefore not applicable in our case. 

The fourth explanation could be that the bulk of the energy is deposited into the deeper, denser chromosphere and the radiation emitted at wavelengths not observed by AIA,  without there being any significant radiation component in the EUV where we measure the DEM. However, in a conductively-heated atmosphere this appears unlikely. How would the bulk of the energy be transported from the corona to this depth, passing through the upper chromospheric and transition region layers without heating those layers and producing detectable emission? For example in the similar situation of the TRACE moss \citep[e.g.][]{1999ApJ...519L..97B,1999ApJ...520L.135F}, the overpressure of the hot coronal SXR loops means that the hydrostatic structure places the million-degree plasma in a fairly dense region of the atmosphere (transition region or upper chromosphere densities) where it radiates strongly. The pressure in the moss is 1-2 dynes~cm$^{-2}$; the pressure in the coronal loop here, from the measured density and temperature is 7 dynes~cm$^{-2}$. Therefore one would expect the million-degree plasma to appear at even deeper, denser locations in this event, and radiate more strongly than does the moss, which is not the case.
\section{Conclusions}
We present observations of a pre-flare hot coronal loop seen by RHESSI at 10~MK that does not seem to be energetically (in terms of thermal conduction) connected to the chromosphere. The pre-flare activity as seen in SXR lightcurves and images and EUV lightcurves and images lasts for $\sim$ 20 minutes before the onset of HXR emission. Energy calculations suggest a conductive energy flux of the order of $10^{10}$ erg s$^{-1}$ cm$^{-2}$ deposited into the chromosphere. However, no chromospheric response is observed. The coronal source appears to be thermally isolated from the chromosphere, yet none of the possible explanations (increased absorption, thermal conduction to elsewhere in the active region or the solar wind, trapping or ``bottling up'' of electrons in the loop, heat deposition to the deep chromosphere with) is applicable for the conditions in the present flare. 

\mdseries
\acknowledgments
Financial support by the European Commission through the FP7 HESPE (FP7-2010-SPACE-263086) is gratefully acknowledged (P.J.A.S., L.F.). This work is also supported by STFC Grant  ST/I001808 (L.F.), by the Swiss National Science Foundation (M.B.), and the International Space Science Institute (ISSI) Bern. We thank S\"am Krucker for helpful discussions, Tereza S. Pinto for synthesizing the NoRH maps used in our analysis, and Gregory Fleishman and Natalia Meshalkina for discussions on SSRT data related to this flare  
\bibliographystyle{aa}
\bibliography{mybib}

\end{document}